%% file: trails.tex
\documentstyle[revtex,eqsecnum]{aps}
\draft

\begin{document}

\begin{title}
Enumeration of self avoiding trails on a square
lattice using a transfer matrix technique.
\end{title}

\author{A R Conway}

\begin{instit}
Department of Mathematics,\\
The  University of Melbourne,\\
Parkville 3052, \\
Australia
\end{instit}

\author{A J Guttmann}
\begin{instit}
Department of Theoretical Physics,\\
Oxford University,\\
1 Keble Road,\\
Oxford OX1 3NP\\
U.K.
\end{instit}


\begin{abstract}

We describe a new algebraic technique, utilising transfer matrices,
for enumerating self-avoiding lattice trails on the square lattice.
We have enumerated trails to 31 steps, and find increased evidence
that trails are in the self-avoiding walk universality class.
Assuming that trails behave like $A \lambda ^n n^{11 \over 32}$, we find
$\lambda = 2.72062 \pm 0.000006$ and $A = 1.272 \pm 0.002$.

\end{abstract}


\def\v#1{\hbox{\boldmath$#1$\unboldmath}}
\def\uv#1{\hat{\v{#1}}}
\def\half{{1\over 2}}
\def\tensor#1{\v{\cal #1}}
\def\animal#1{\tensor{#1}}

\def\idrevtex#1{#1}
\def\idlatex#1{}

\def\figcaption#1{\idlatex{\caption{#1}} \idrevtex{#1} }

\def\figfig#1#2#3{\begin{figure}
\figcaption{#1}\idlatex {\vspace{8cm}
\includegraphics{#2} }
\label{#3} \end{figure}
}



\section{History}

Over the years, the study of the trails problem has provided an interesting
counterpoint to the corresponding SAW problem. While self avoiding walks are
connected open non-intersecting paths on a lattice, and hence no site or bond
may be visited more than once, lattice trails are open paths on a lattice
which may re-visit sites, but not bonds. Thus SAWs are a proper subset of
trails. First seriously studied by
Malakis \cite{trails:first},
a number of exact and numerical results were obtained
by Guttmann \cite{trails:exact,trails:num}.

It has
been shown by Hammersley \cite{saw:ham} that SAWs have a connective
constant: that is some value $\mu$ such that if there are $c_n$
SAWs of length $n$, then $\log \mu= lim_{n\to\infty} \log(c_n) / n $
exists and is finite and non-zero.
Later, Hammersley and Welsh proved that
$c_n=\mu^n \exp(O(\sqrt{n}))$.

These results were carried over to trails by
Guttmann\cite{trails:exact}.
If $t_n$ is the number of trails of length $n$,
and $\lambda$ denotes the connective constant for trails then
$c_n\leq t_n$ and $\mu \leq \lambda$.
It was also shown\cite{trails:exact} that the critical
behaviour is in the saw universality class for trails on the honeycomb
lattice, no such proof has been found for the square lattice case. The
earlier series were found to be rather poorly
 converged compared to saw series of
similar length, with the exponent of the trails generating function being
 $\gamma \approx 1.40$, compared to the saw result $\gamma = {{43}\over{32}}
= 1.34375$. The connective constant was estimated as
$\lambda = 2.7215 \pm 0.002$.
This poor convergence prompted Guttmann and Osborn \cite{trails:montecarlo}
to carry out a
Monte-Carlo study, using the Berretti-Sokal \cite{saw:montecarlo}
algorithm, using walks
up to 200 steps. They found $\gamma = 2.7205 \pm 0.0016$
and $\gamma = 1.348 \pm 0.11$. A biased estimate of the connective constant,
assuming $\gamma = 1.34375$, gave the critical point estimate $\lambda =
2.72059 \pm 0.0008$. Recently Lim and Meirovitch \cite{trails:scan}
used an entirely different
Monte Carlo algorithm, the scanning simulation method. They obtained
the estimates $\lambda = 2.72058 \pm 0.00020$ and $\gamma = 1.350 \pm 0.012$.

In this work we report a substantial extension of the series expansion
of the generating function for square lattice trails.
The finite-lattice method plus transfer matrices described here allows
31 terms to be obtained on a work station (an IBM 6000/530 with 256MB of
memory). The method is described below.
The complexity of our algorithm is in fact worse than exponential,
compared to $\lambda^{n}$
required by a conventional enumeration algorithm, where n is the maximum
number of steps.
However for intermediate values of length, say 50-100 steps, it is
in fact substantially faster. The detailed performance is discussed in
Section \ref{algcomplexity}

\def\picP{ \parbox{51pt}{\input{P}}}
\def\picQ{ \parbox{51pt}{\input{Q}}}
\def\picR{ \parbox{29pt}{\input{R}}}
\def\picS{ \parbox{51pt}{\input{S}}}
\def\picT{ \parbox{51pt}{\input{T}}}
\def\picV{ \parbox{73pt}{\input{V}}}
\def\picX{ \parbox{51pt}{\input{X}}}
\def\picwzX{ \parbox{51pt}{\input{wzX}}}
\def\picred{ \parbox{95pt}{\input{red}}}

\section{Algorithm}

The algorithm for enumerating self avoiding trails is very similar to
the algorithm for enumerating self avoiding walks described in
\cite{saw:saw39}. We will summarise this process, expanding
and pointing out the
differences between walks and trails where appropriate. Where there
is no difference, the word {\it paths} is used to denote either
walks or trails.

\subsection{Introduction}

The basis of this method is the transfer matrix technique
on a finite lattice. This enables
one to count the total number of paths on a square lattice (or other
type of lattice, with the appropriate modifications).  We shall firstly
discuss paths that can fit into a finite lattice.

\label{algintro}

The fundamental problem with enumerating self avoiding paths
that the self avoiding constraint is non-local. One can't just say, for
instance ``There are $x$ number of ways of getting from $(0,0)$ to
$(a,b)$ and $y$ number of ways to get from $(a,b)$ to $(c,d)$, so there must
then be $xy$ ways to get from $(0,0)$ to $(c,d)$.'' However, if we
{\it could} do something like this, it could save
a great deal of time, since
$xy$ is typically much larger than $x+y$...that is, it is faster to
count $x$ steps followed by $y$ steps than it is to count $xy$ steps.

If one draws a boundary line through the (finite) lattice, one notices that
the self avoiding constraint works independently on both sides. This
means that it would be possible to work out, for all the possible
boundary conditions, how many patterns to the left and right
of the boundary
there are that give those boundary conditions.
A boundary condition is the set of bonds cut by the boundary line,
plus a description of their interconnectedness.
Thus, one can consider
the two partial paths given in figure~\ref{exampbound} as behaving in exactly
the same manner, if all further growth takes place on the right of
the boundary. The number of partial paths to the left and right with a given
boundary can then be multiplied and summed over all boundaries to give
the required number of paths.

Note that we usually do not want the total number of paths on a certain
sized lattice, but rather the number of paths of a certain length
on that lattice.

To cope with this, instead of just counting the number of partial paths to
either side of the boundary, one can count the number of partial paths
of $n$ steps, $g_n$, and then make a generating function
$G(x)=\sum_{n=0}^\infty g_n x^n$. Then, when one multiplies the
generating functions for either side of the boundary, one ends up
with a total generating function, from which the number of paths of
the appropriate length can be easily extracted.

However, this leaves the tasks of actually counting those paths,
and of matching them up. This task can be simplified by noticing
that we could add in a second boundary (figure~\ref{twobound})
separating just a single site.

Now, we have three independent areas. The matching process is however
not much more complicated. First, one works out the generating functions
for all possible boundary conditions on the left.
Secondly, for each of the boundary conditions on the left,
one works out which new paths can be created by adding this new point.
This will create zero or more possible new paths to the left of the
second line, with perhaps different generating functions. When two
or more different ``first boundary'' conditions create the same
``second boundary'' condition, the generating functions should be added.
An example of the new boundary conditions created is shown in
figure~\ref{exampnew}

This process can be represented as matrix multiplication. We start off with a
column vector of generating functions, where each element corresponds
to a particular (first) boundary condition. We then perform a linear
transformation of this column vector to another column vector containing
a generating function for each (second) boundary condition. This is the same
as multiplying on the left by a rectangular matrix. This is the reason
for the name ``transfer matrix technique''.
Note that the column matrices are very long, so this rectangular matrix
will be exceedingly large. However, there is no need to store this matrix
as the matrix is very sparse, and elements can be calculated on demand.

This process can add one extra point. But there is no reason that this
process can not be continued to add a second point... or a third...
or the rest of the lattice, one site at a time. Once the end
of the lattice is reached, the boundary conditions are very simple to
match: nothing is allowed to cross the end of the lattice.
Similarly, there is no reason why one cannot start at the very beginning
of the lattice.
One then begins with
only one initial boundary
condition --- no connections, and the trivial generating function $1$.
Thus one can generate all the possible paths using only the transfer matrix
techniques. This means one never has to explicitly count paths. This
process is potentially significantly faster than counting paths
individually.

\subsection{Defining Boundary conditions}

The boundary specification is called a ``signature'' and is based on
a series of numbers, one for each of
the bonds crossed by the boundary line. For a lattice of
width $W$ bonds, there will be $W+1$ (vertical line)
or $W+2$ (vertical line with a kink) of these bonds.

\label{algbound}

Each bond crossed may be
characterised by one of three possibilities. Firstly, it may
be unoccupied, in which case it is easy to specify. Assign it
the number `0'. Secondly, the bond may be
occupied, and lead to a dangling end. That is,
the pathlet connected to the bond ends somewhere to the left
of the boundary. This is also easy to specify --- assign it
the number `1'. Note that these assignments are arbitrary, there
is no inherent meaning in this code. The third possibility is
that the bond is occupied, and is connected (via some route
to the left of the boundary) to some other bond on the boundary.
In order to fully specify this, one must somehow uniquely
define which bond it is connected to.

The arrangement of possible interconnections is severely limited in
the self avoiding walk case, as the pathlets cannot cross. This enables
a very efficient encoding: If one labels the `top' of a pathlet with
a `2' and the bottom with a `3', then this will uniquely specify
the way the bonds are connected. For, if there is a `2' at some
point in the signature, one can find the corresponding `3' by
moving down the signature until the next `3' is found, subject
to the condition that every time a `2' is crossed, one must ignore
an extra `3'. So for instance, in the signature ``232233'',
the first `2' matches the first `3', whilst the second `2'
matches the last `3'. A computer can then store each code number for
each bond as two binary bits, so for $W\leq 14$
the whole signature fits nicely into a 32 bit binary word,
which is very convenient for current computers.

For self avoiding trails, we do not have this nice restriction.
Self avoiding trails can cross themselves at a site, as it
is only the bonds that have to avoid one another. This means that
one cannot get away with a clever encoding of just two
symbols. Indeed, no finite number of symbols will do for all
values of $W$, as will be presently shown. This leaves the
explicit option, where the code number for each bond specifies the index
of the bond to which it is attached (plus 1). The index of a bond
is the bond's position on the boundary --- $1$ up to $W+2$ inclusive.
The addition of one is to prevent mistaking a dangling end with a connection
to the first bond. This means each bond will have a number from
0 to $W+3$ associated with it. If we restrict $W\leq 12$ in a computer
program, then each bond fits into 4 bits, and the total signature
is 56 bits, which is a little more clumsy to deal with, but not
difficult.

Note that the specific numbers mentioned above for restrictions
on $W$ are in no way restrictions on the algorithm, just on a
particular programming implementation. We only used $W=7$
anyway, due to finite computer resources.

Paradoxically, the ``implicit'' coding of the self avoiding
walk boundaries is significantly easier for a computer
program to deal with than the
``explicit'' encoding for the self avoiding trails. This is due
to the fact that most operations deal with local changes, when adding
a site. These local changes are easy to implement with the implicit
coding.

Note that in \cite{saw:saw39} we used a permutation of the numbering
system described above for SAWs: we used `3' as the dangling end
marker, and `1' and `2' for the loop ends. We have changed notation here
for consistency.

The coding of the signatures are actually of vital
importance, as the total time and memory requirements of the algorithm are
polynomials in $W$ times the number of different signatures ($W^2$ for
space, and $W^4$ for time), as discussed in section \ref{algcomplexity}

For self avoiding walks, an upper bound on the number of different
signatures is obvious: $4^{W+2}$, as there are $4$ different possibilities
for each bond, and $W+2$ possible bonds. Actually, there are significantly
fewer than this since not all combinations are possible: one can't have
more than two floating ends (as it would be impossible to make
them into a connected walk), and one can't end a loop with a $3$ before
starting it with a $2$. It turns out that the number of possible
signatures grows like a polynomial in $W$ times $3^W$.

For self avoiding trails, the situation is again worse.
The number of possible boundary conditions can be evaluated
exactly in a straight forward manner. Let $B_n$ be the number of
boundary conditions for a strip of width $n$, and $L_n$ the
number of boundary conditions {\it excluding ``dangling'' paths}.
That is, all of the $n+2$ bonds on a boundary condition counted
in $L_n$ must either be unused bonds, or may be connected to
another bond on the boundary. In \cite{saw:saw39}
these were marked by a `3'. A formula for $B_n$ in
terms of $L_n$ is easy to obtain: any boundary counted in $B_n$
may have no dangling ends (giving a term $L_n$), or it   may have
one dangling end in any of $n+2$ places (giving a term of
$(n+2)L_{n-1}$), or it may have two dangling ends giving a term
of $(n+2)(n+1)L_{n-2}$. Thus we have
\begin{equation}
B_n=L_n+(n+2)L_{n-1}+(n+2)(n+1)L_{n-2}
\label{defB}
\end{equation}
Now to work on an equation for $L_n$: The first bond may be
unoccupied, giving a term of $L_{n-1}$, or it may be
connected to one of $n+1$ other bonds, giving a term of
$(n+1)L_{n-2}$. Thus
\begin{equation}
L_n=L_{n-1}+(n+1)L_{n-2}
\label{defL}
\end{equation}
Initial conditions are $L_{-1}=1$, $L_{-2}=1$ and $L_{-3}=0$.

These are worrying equations as they grow faster than exponentially
due to the $(n+1)L_{n-2}$ term in equation (\ref{defL}). This faster
than exponential growth is the reason why no finite set of symbols
could cope with encoding the connections for all values of $W$.

Actual values
are given in table~\ref{BLtable}, along with the number $s_n$
(taken from \cite{saw:saw39}) of
possible boundary conditions for self avoiding walks.

\subsection{Irreducible Components}

Directly using the transfer matrix method is not as much of a saving
as could be expected, due to the very large number of vectors. If we want
to count all paths up to a maximum length of $2n+1$, then at first
it looks as though a square $2n+1$ wide is needed to cope with a perfectly
vertical path. However, it is possible to use the symmetry relation between
the horizontal and vertical axes, so that only paths up to a width of
$n$ need to be calculated: for paths of width  $n+1$ or greater, we can
say that they must have height $n$ or less, and thus their mirror images
will have already been counted. More formally, if $G_{ij}$ is
the number of paths with $i$ horizontal and $j$ vertical steps, then
$G_{ij}=G_{ji}$ and thus if we know $G_{ij}$ for $i\leq 2n+1$ and
$j\leq n$, we really know $G_{ij}$ for all $i+j\leq 2n+1$. That is,
we know the total number of paths of length up to $2n+1$ steps.
This means that we could work with strips of width $n$, length $2n+1$ and
obtain coefficients up to and including $2n+1$.

There is still a further improvement. Suppose that we break up all the
paths (of vertical steps $\leq n=2M+1$) into two classes:

{\bf Irreducible} paths have no place where a horizontal line
	could be drawn across the lattice intersecting exactly
	one vertical bond. As there are a maximum of $2M+1$ vertical
	bonds in the path, and there must be at least two vertical bonds
	for each unit of width (to satisfy the irreducibility
	definition),
	these paths must all fit into a strip of width $M$ bonds.

{\bf Reducible} paths have at least one
	place where the horizontal line
	can be drawn, intersecting just one bond of the path.
	These paths have the nice
	property that the self avoiding constraint will act independently
	both above and below this line. All that is needed is to calculate
	the number of self avoiding paths above and below independently.
	This is a smaller problem, and indeed, can be further split up,
	until the entire path can be considered to be made up
	of an irreducible ``top'' section, then one or more sections
	composed of a vertical bond and an irreducible ``middle'' section,
	then finally a vertical bond and an irreducible ``bottom'' section.
	As all of these irreducible subsections will have fewer than
	$n$ steps, they will fit into a strip of width $M$.

All in all, these two optimisations allow calculations on a strip of width
$M$ bonds to provide the number of paths with widths up to $n=2M+1$ and
thus paths with total number of bonds up to $2n+1=4M+3$. Since the number
of partial generating functions rises exponentially with strip width,
these two optimisations reduce the complexity of the problem enormously.

However, it makes the counting task a little more difficult: we have to
extract these ``top'', ``middle'' and ``bottom'' sections individually.
To facilitate this, the irreducible paths can be named as described
below, based upon their starting and end points. Note that a distinction
is made here between paths and routes. A {\bf path} has a specific
starting point: a {\bf route} does not. This means that there are exactly
half as many routes as paths.

\def\descbox#1{\parbox{6cm}{#1}}

$$
\begin{tabular}{|c|c|c|c|}
\hline
\descbox{Description} & \hbox{Picture} & Name & \parbox{4cm}
   {Lowest power in w for width M} \\
\hline \hline
& & & \\
\descbox{Path with no vertical bonds}         &  \parbox{51pt}{\input{P}} & P & n/a\\
& & & \\
\descbox{Route with two middle ends}          &  \parbox{51pt}{\input{Q}} & Q & $w^{2M}$ \\
& & & \\
\descbox{Route with two top ends}             &  \parbox{29pt}{\input{R}} & R & $w^{2M}$ \\
& & & \\
\descbox{Route with one top, one middle end}  &  \parbox{51pt}{\input{S}} & S & $w^{2M+1}$ \\
& & & \\
\descbox{Route with one top, one bottom end}  &  \parbox{51pt}{\input{T}} & T & $w^{3M}$ \\
& & & \\
\hline
\end{tabular}
$$

Note that routes with two bottom ends are not included, as they are
the same (in number and shape) as $R$, and similarly routes with one
bottom end and one middle end are not given a name as they are covered
by $S$. Note that all the routes above are irreducible.

The name in this table is the name of the generating functions associated
with that variable in this paper. There are six generating functions
associated with each letter in this paper, as per the following
pattern:
\begin{itemize}
\item	$Q(u,w)$ is the generating function for irreducible
	routes of the required shape with the power of $u$ giving
	the number of horizontal bonds, and $w$ representing
	vertical bonds.
\item   $Q_W(u,w)$ is the same, except only for those irreducible
	routes of width exactly $W$.
\item	$Q^*(u,w)$ is the generating function for {\it all} (i.e. both
	reducible and irreducible) routes of the required shape.
\item	$Q^*_W(u,w)$ is the same, except for all
	routes with width exactly $W$.
\item   $Q(u,w,z)$ is the generating function for irreducible
	routes of the required shape with the power of $u$ giving
	the number of horizontal bonds, $w$ representing
	vertical bonds, and $z$ the total width.
\item   $Q^*(u,w,z)$ is the generating function for all
	routes of the required shape with the power of $u$ giving
	the number of horizontal bonds, $w$ representing
	vertical bonds, and $z$ the total width.
\end{itemize}
Note that the same terminology applies to variables other than $Q$, with
{\it routes} changed to {paths} where appropriate. The three variable
generating function is the most general: the width $W$ generating functions
can be extracted from the appropriate power of $z$, and the
generating functions in two variables can be produced from
the functions in three variables by setting $z=1$.  That is,
$$Q(u,w)=Q(u,w,1)$$
$$Q(u,w,z)=\sum_{n=0}^{\infty} z^n Q_n(u,w)$$

Note that if a path is on a strip the width of which is
too small for the definition to
make sense, then the corresponding generating function is zero: i.e.
$Q_0$, $Q_1$, $R_0$, $S_0$, $S_1$, and $T_0$ are all zero.

Of these five functions, $P$ is easy to determine. There is one horizontal
path of length zero, and two paths of every other length
(one in each direction). Thus
$$P(u,w,z)=1+2u+2u^2+2u^3+....={{1+u}\over{1-u}}$$

Now define another variable, $X$. This will represent the total number
of irreducible middle sections. That is, the number of
ways of going from a point at the bottom
of an irreducible section to a point on the top. Note that every element of
$T$ can be considered as a path {\it restricted so as to not go below the
starting point}. Thus, $T$ copes with all the parts of $X$ of width
at least one. For the zero width case, we just want paths from one point
on a line to another point $P$. Thus $$X=T+P.$$ This is the reason
for defining $P$ to be paths, whilst $Q$, $R$, $S$, and $T$ are routes.

This is a typical $X$:  \parbox{51pt}{\input{X}}, and this is the corresponding $wzX$:  \parbox{51pt}{\input{wzX}}

$X$ refers to just a single irreducible middle section. This can be extended
to an arbitrary middle section by noting that a ``middle section'' can be
formed from either a vertical bond ($wz$), or two vertical bonds with
an $X$ in between, ($wzXwx$), or any number of extra $wzX$ terms.
Define a new variable $V$ to be a total (reducible) ``middle section'',
then
\begin{equation}
V=wz\left(1+wzX+(wzX)^2+(wzX)^3+...\right)={{wz}\over{1-wzX}}.
\label{eqnV}
\end{equation}
\hbox {
\parbox{12cm}{
A $V$ can be considered to be a generalisation of a vertical bond: It is
a reducible path without either the top or bottom irreducible components. A
typical element of $V$ is shown at the right. The arrows indicate that
the $V$ is intended to be used as {\it part} of a path, not as something
in its own right.
}
 \parbox{73pt}{\input{V}}
}

Note that the top and bottom of a $V$ are {\it always} vertical bonds, so a
$V$ can attach to {\it any} irreducible component which has an end at its
top or bottom. This can be a $P$, an $R$, an $S$ or a $T$. Note that
the $R$ has two ends to which connections can be made,
so we must count it twice.
$P$ is not counted twice since it is a path, not a route. Define
the generating function of end components, $E$ as
$$E=P+2R+S+T.$$

Now all the reducible routes can be calculated. Each consists of
one end piece, $E$, a joint $V$ and another end piece $E$. Thus
reducible routes are $EVE$. Irreducible routes (with some vertical
component, i.e. not $P$) are $Q+2R+2S+T$.
$R$ and $S$ are counted twice to allow for routes with two bottom ends or
one bottom and one middle end respectively. To get the total number of
paths then, we take the number of paths with no vertical component,
$P$, and add in twice the number of routes with vertical
components. This gives
$$C=P+2(Q+2R+2S+T+EVE)$$ as the total number of paths.

\hbox {
\parbox{10cm}{This is a typical reducible path, made up from a $TVP$, where
the $V$ in this case is $wzPwzTwz$.}
 \parbox{95pt}{\input{red}}
}

\subsection{Obtaining the irreducible components}

So far only $P(u,w,z)$ is known. In order to calculate the number of self
avoiding paths up to length $4M+3$,
$Q(u,w)$, $R(u,w)$, $S(u,w)$ and $T(u,w)$ must be known
accurate to $u^{4M+3}$ and to $w^{2M+1}$.

Suppose that it were possible to obtain the starred polynomials
$Q^*$, $R^*$, $S^*$ and $T^*$ as functions of three variables. Then
$R=R^*$, as all paths starting from the top and ending at the top are
irreducible.

Calculating the others is a little more difficult. Consider the generalisation
of $X$ to $X^*$. $X^*$ will be equal to the sum of the irreducible
parts $X$, plus reducible paths starting at the bottom and ending at the top.
These are expressible as $XVX$, so we have $X^*=X+XVX$. Using
equation~\ref{eqnV}, this can be inverted to give
\begin{equation}
X={{X^*}\over{1+wzX^*}}
\label{eqnXXstar}
\end{equation}
which can be expanded in a formal binomial series to give
$$X=X^* \left( 1 - wzX^* + w^2 z^2 X^{*2} - ... \right) $$
If $X^*$ is known to some order in $u$ and $w$ for powers up to $z^M$, then
$X$ can be determined to the same order. Since $X$ is made up
of $P$ (which is zero for widths other than 0), and $T$, which has the
lowest power of $w$ being three times the power of $z$, order is preserved
up to $w^{3M+2}$ and to the original order in $u$. Thus, if $X^*$ is known
to $u^{4M+3}$ and $v^{2M+1}$, this is preserved in the
calculation of $X$.
So, by using the third variable, one can go from $X^*$ to $X$, and thence
$T$. Without using the third variable $z$, the generating function
$X^*$ would only be correct to terms of order $w^M$ rather than
$w^{2M+1}$.

Similarly, if we define $Y=2R+S$ (connections at the bottom,
but not the top), then $Y^*=2R^*+S^*=2R+S+XVY=Y(1+XV)$, so
there is an expression for $Y$ similar to equation (\ref{eqnXXstar}),
$$Y={{Y^*}\over{1+XV}}$$
One can then obtain $Y$ and thence $S$ from $Y^*$ and hence $S^*$, in a
manner similar to that used to obtain $T$ from $T^*$ via $X$ and $X^*$.

Lastly, $Q^*=Q+YVY$ so $$Q=Q^*-YVY$$
and $Q$ can also be obtained in a similar manner.

\label{usered}

This means that all the irreducible components can be obtained from reducible
components given the full three variable information, and accuracy to
\begin{itemize}
\item $M$ in $z$ (i.e. to width $M$)
\item $2M+1$ in $w$
\item $4M+3$ in $u$.
\end{itemize}

\subsection{Obtaining reducible components}

Suppose that we could count all the paths on a certain finite lattice
with constraints upon where the paths can start or end. Define the
generating function in variables $u$ to order $4M+3$ and
$w$ to order $2M+1$ for paths on a strip of width $K$ as $G_K(a,b,c)$,
where $a$, $b$, and $c$ are $+$ or $-$ depending upon whether
one can start or end paths on the top of the strip, the bottom
of the strip, and/or the middle of the strip respectively. Ensure
that all paths included in these generating functions start flush at
the left of the lattice so that we do not need to worry about
uniqueness in the horizontal direction.

\label{getred}

Now, by considering how the walks that fit into the strip can be made up
of the reducible functions defined above, the latter can be defined as
an invertible linear combination of the former. One inverts this
relation and gets the reducible components needed in section \ref{usered}
from the $G_K(+,-,-)$, $G_K(+,+,-)$, $G_K(-,-,+)$ and $G_K(+,-,+)$,
for $K$ from $0$ to $M$.

These relations are (as taken from \cite{saw:saw39})
\begin{eqnarray*}
 R_{m}& =& G_{m}(+,-,-) - G_{m-1}(+,-,-) \\
 Q^{*}_{m}& =& G_{m}(-,-,+) - G_{m-1}(+,-,+) - \sum_{n=1}^{m-1}
    \left(Q_n^*+R_n+S_n^*\right) \\
 S^{*}_{m} &=& G_{m}(+,-,+) - G_{m-1}(+,-,+) - G_{m}(+,-,-) -
    G_{m-1}(+,-,-) \\
 && \quad - {{P-1}\over 2} -Q_m^* - \sum_{n=1}^{m-1}
    \left(Q_n^*+2*S_n^*+T_m^*\right) \\
 T^{*}_{m}& =& G_{m}(+,+,-) - 2G_{m}(+,-,-) \\
\end{eqnarray*}

\subsection{Counting paths on strips}

The transfer matrix technique can be used to obtain the
generating functions $G_K$
that are needed in section \ref{getred}.

Suppose we are working on a lattice of width $W$ and length $4W+3$.
As mentioned before, one starts with one partial generating function
(boundary to the left of the entire lattice, no bonds used, generating
function 1). Then add on sites as described in the next paragraph
one at a time, working along the matrix column by column. At each
site one stores for each valid signature the partial generating
function. After processing the first column, one can remove the
signature with no bonds occupied, as any animal based upon this
signature will not lie flush against the left of the lattice, and
by removing it we satisfy the horizontal uniqueness criterion.

To process a site, one cycles through all the stored signatures,
processing each individually, creating a new set of signatures. Note that
two or more signatures may produce the same signature after processing.
In this case the partial generating functions for these two signatures
should be added.

All that is left is to describe exactly what to do when each site is
added for a particular signature.
The site that is being added will have two bonds coming
in (to the left of the new boundary), and another two bonds leaving
(to the right).

One must firstly see if the walk
can be finished at this point, and if so, add in the partial generating
function to a total generating function which will give the final
$G_K$ once all sites have been processed.
In order to be able to accumulate a partial walk, two conditions must
be satisfied. Firstly, there must be no occupied bonds
in the signature other than those coming into the bond being processed.
Secondly, one of the three following conditions must hold:
\begin{itemize}
\item There must be a single dangling end coming in to
the site being processed (type `1' in the signature coding), and it
is valid to start or stop a path at this point (determined by
the $+$ or $-$ parameters in the particular $G_K$ being
computed.
\item Or there may be two dangling ends that connect at this site.
\item Or  (only in the case of trails) there may be a loop
completed at this site and it is valid to start or stop
a path at this point.
\end{itemize}

We will first discuss the possibilities for the new
signatures if one cannot start or stop a path at the site
being processed.

If one is counting walks, and there is only one bond going
into the site, then that one bond must emerge from either of the two
bonds coming out of the site. This gives two new signatures, one
with the old generating function multiplied by $w$ (emerging
vertically), and one with the old generating function multiplied by
$u$ (emerging horizontally). In future we will not mention these
multiplications.

Again for walks, one may have both bonds entering the site
occupied. In this case neither output bond may be occupied, as one
can't have more than two occupied bonds touching a site for
self avoiding walks. What happens depends upon the specific case.
If the two bonds are attached together, then a loop has been
formed which is illegal, so no signatures are generated.
If the two bonds are dangling ends, then attaching them would
make an entire dangling path, which is not allowed. In the remaining
cases, one does produce a new valid signature, and one must adjust
the coding for the bond(s) in the signature to which the just
processed bonds were attached.

Again for walks, if there are no bonds coming in, then there are
two possibilities:
no bonds coming out, or a new path being started
at this point -- that is two bonds coming out and connected to each
other.

Further possibilities exist if one can stop or start
from the site being processed.

For walks with no bonds coming in, one can now have one dangling
end coming out either of the two
outgoing bonds. With one bond coming in
which is not a dangling bond, the pathlet it belongs to can be terminated
at this site, and the bond to which it is attached elsewhere in
the signature becomes a dangling bond. Note that each of these steps
increases the number of dangling bonds in the signature, and one
must check that the total number of dangling bonds does not
exceed two, as this would mean that any path one tries to construct
must have at least three ends!

These are summarised in table 2 of \cite{saw:saw39}.

For trails, the situation is significantly more complicated, as
bond loops and crossings are allowed, but the basic idea remains
the same.

First, consider what can be done without starting or stopping.

The same possibilities as in the walks case (without
stopping or starting at the site being processed) hold, with some
extra possibilities when there are two bonds coming in. Firstly,
both bonds could ``bounce'' and come out as two bonds with the
same connections. Secondly, they could cross, and come out as
two bonds with interchanged connections. Thirdly, if the two bonds
coming in meet, and in the walks case would have produced no
bonds coming out, one may also have two new connected bonds coming out,
as occurred in the walks case when no bonds went in.

If one is allowed to start or stop at the site being processed,
things get much more complicated. The actions can best be
described by two stages.

In the first stage, associated with terminating incoming
pathlets, one forms all the possibilities already
described, and adds in the following possibilities:
\begin{itemize}
\item
For one bond entering which is not an dangling end, the pathlet
may be terminated at this site, and the other end of the pathlet
converted to a dangling end (as was done for walks). No occupied bonds
emerge.
\item
For two bonds entering, one a dangling end, and the other a pathlet,
the pathlet may terminate (making the other end of the pathlet a
dangling bond) and the dangling end can continue from either of the
two new bonds.
\item
For two ends of the same pathlet entering,
one end may terminate at the current site, and the
other end (now a dangling end)
 may take either of the two new bonds. As either end of the
pathlet may terminate, there are four new signatures produced.
\item
For two ends of different pathlets entering, there are the
same four possibilities as above, except that this time it
is a pathlet leaving, not a dangling end, and some other bond
in the signature will become a dangling end. A fifth
possibility is for both incoming pathlets to terminate, producing
two dangling ends elsewhere in the signature and no bonds
coming out.
\end{itemize}
In the first and last case above, there is the possibility of
no bonds coming out. Again, one can add a new two bond loop
in both cases as in the walks case when no bonds went in.

The second stage is associated with adding dangling ends
at the leaving stage. If any of the signatures formed from the
first stage have either or both of the outgoing bonds
unoccupied, either or both may be filled with dangling ends.

\label{calcG}

Of course, when forming new dangling ends, one must remember the
constraint that the total number of dangling ends in the
signature may not exceed two.

\subsection{Algorithm complexity}

One now has all the ingredients for the algorithm. One uses the
transfer matrix technique to get all the $G_K$ terms for $K$ going
up to some value $W$ (\ref{calcG}), then obtain the reducible
generating functions (\ref{getred}) and thus obtain the irreducible
generating functions and final answer (\ref{usered}).

\label{algcomplexity}

Of these three stages, the first (\ref{calcG})
is exceedingly time and memory consuming, whilst the second
(\ref{getred} and \ref{usered})
is fast (polynomial in $W$ time) and uses little memory.

Since the first stage is the bottleneck, we shall discuss it exclusively
in terms of complexity.

The total memory required will be bounded by the number of possible
boundary conditions, multiplied by the total space per generating
function (proportional to $W^2$), multiplied by two, since one may need
to store both the incoming and outgoing partial generating function. In
practice, this last factor is nowhere near as high as two, since as soon
as a signature has been fully processed, the data associated with it
may be discarded.

The total time required is proportional to the total amount
of memory that needs to be processed (as above) times the number of
sites that have to be processed (proportional to $W^2$), times the
average number of new signatures per old signature. This last factor
is pretty much independent of $W$. For trails it is significantly larger
than walks.

The basic result is that the time and memory requirements are a
small polynomial times the number of boundary conditions. The
number of boundary conditions is therefore the most significant
factor in the complexity of this algorithm.

For self avoiding walks, the number of boundary conditions grows
like a polynomial in $W$ times $3^W$. Thus the
dominant complexity of
this method for self avoiding walks is $3^{n\over 4}$, where $n$ is
the number of steps required. This comes from the fact that
$n=4W+3$. The alternative, direct enumeration, grows like $\lambda^n$,
where $\lambda$ is the connective constant for self avoiding walks.
Note that $\lambda$ is significantly greater than $3^{1\over 4}$
(approximately twice $3^{1\over 4}$ in fact),
so this algorithm is exponentially faster than direct enumeration.

For trails, the situation is not as good. The analysis in section
\ref{algbound} shows that the number of boundary conditions grows
faster than exponentially. Thus, for very long trails,
direct enumeration will be a more efficient algorithm! However,
consulting table \ref{BLtable} shows that trails are not all that much
worse than walks for small values of $W$. So for small
values of $W$, this transfer matrix method is actually more
efficient than direct enumeration. Fortunately,
the values of $W$ for which this algorithm is faster
than directed enumeration are such that this algorithm
is faster for $n$ at least 50, which is far beyond the
capacity of current computers.

This algorithm is also amenable to parallelisation in the same manner as
the self avoiding walk algorithm described in \cite{saw:saw39}.

This algorithm was implemented in a C program using modular
arithmetic, and was used to obtain trails of up to 31 steps.
They are given in table \ref{tabenum}.

\section {Analysis of series}

The method of analysis used is based on first and second order
differential
approximants. It was used in previous
papers \cite{alg:critwalk,saw:saw29,saw:saw39}
in which the related saw problem was studied, and is described
in detail in \cite{alg:diff}. In summary, we construct near-diagonal
inhomogeneous differential approximants, with the
degree of the inhomogeneous polynomial
increasing from 1 to 8 in steps of 1. For first order approximants
(K=1),
 12 approximants are constructed that utilise a given number of series
coefficients, N. Rejecting occasional defective approximants, we form
the mean of the estimates of the critical point and critical exponent
for fixed order of the series, N. The error is assumed to be two
standard deviations. A simple statistical procedure combines the
estimates
for different values of N by weighting them according to the error,
with the estimate with the smallest error having the greatest weight.
As the error tends to decrease with the number of terms used in the
approximant, this procedure effectively weights approximants
derived from a larger number of terms
more heavily.

For second order approximants (K=2),  8 distinct
approximants are constructed for each value of N.
We find that as the number of series terms increases, the estimate
of the critical exponent decreases. We show below that this is due to
rather strong correction-to-scaling terms, much stronger than for
the saw case.
Because of this, the estimates we quote below should be treated as
over estimates of the exponent and critical point.
        $$x_c = 0.367597 \pm 0.00002
\quad		\gamma = 1.352  \pm 0.01	\quad	(K=1)$$
	$$x_c = 0.3676 \pm 0.0001
       \quad \gamma = 1.348  \pm 0.008        \quad     (K=2)$$

These results provide some support
for the view that the trails are in the saw universality class.  The
critical point estimate can be refined if we assume that
 $\gamma = 1.34375$ exactly, which is the saw value. To refine the
estimate of the critical point, linear regression is used. There is a
strong correlation between estimates of the critical point and critical
exponent. This is quantified by linear regression, and in this way the
biased estimates (biased at $\gamma = 43/32$)
are obtained.

We find
	$$x_c = 0.367564  \pm 0.000008 \quad (K=1)$$
	$$x_c = 0.367562 \pm 0.000007\quad  (K=2)$$

These are combined to give our best estimate for the connective constant
$\lambda = 1/x_c = 2.72062 \pm 0.00006$, which is in agreement with previous
estimates, but rather more accurate than any previous estimate.

The much slower rate of convergence of the trails series critical point
estimates compared to the corresponding saw estimates is
presumably due to stronger
``correction-to-scaling'' terms. We have investigated this possibility
using three different methods.
Firstly, we used the method of Baker and Hunter
\cite{saw:transform} which transforms the series
so that poles of the Pad\'e approximants to
the transformed series furnish estimates of the reciprocals of the
exponents. However we found that the singularity on the negative real axis
at $-x_c$ masked the presence of any
confluent singularity at $x_c$. Accordingly,
we split the series in two, treating the odd and even subsequences as
independent series. In this way, we found
exponents with the values $\approx 1.35$
and $\approx 1.0$ from the even sub-sequence.
The smaller exponent was not well
identified however. This implies a correction-to-scaling
exponent of $\approx 0.35$.
The odd subsequence gave no evidence of any
exponent apart from the leading one.

The next method we used was the method of Adler et al.
\cite{saw:scaling}, in which
a correction-to-scaling exponent is assumed,
and then a transformation is applied
which maps this non-analytic correction term to an analytic correction term.
Pad\'e analysis of the transformed series should
then give the correct leading exponent.
We tried various values of the correction to
scaling exponent, and found that a
value around $0.75$ resulted in a series which gave the correct critical
exponent of $\gamma = 1.34375$.

The third method  is the same as that used in our recent study of saws
\cite{saw:saw39}. In that method we {\it assume} the correction to scaling
exponent, and fit the series coefficients to the assumed form. The fit
is judged reasonable if the sequences of amplitude estimates appear to
converge well. This is not a particularly sensitive method, but is
useful in that it does provide amplitude estimates as well as.
From the two values
of the correction-to-scaling exponent found above, we tried an intermediate
value of $0.5$. Given that the s.a.w. exponent appears to be $1.5$, this
seemed a reasonable thing to try.
As well as the correction-to-scaling term,
there is another singularity on the
negative real axis. For saws, Guttmann and Whittington
\cite{saw:negroot} showed that this
was at $x = -x_c$.
That proof applies {\it mutatis mutandis} to trails.
We assume that universality of
exponents applies to non-physical singularities
also - a result supported by our series analysis. Then the singularity on
the negative real axis will also have the same exponent as the energy at the
physical singularity - as for saws - and so we expect the generating function
for trails to behave like
\begin{eqnarray*}
T(x) &=& \Sigma t_nx^n \sim A(x)(1-\lambda x)^{-43/32}[1 +
           B(x)(1-\lambda x)^{\Delta} ... ]  \\
		&& \quad + D(x)(1+\lambda x)^{-1/2}.
\end{eqnarray*}

 The exponent for the singularity on the negative real axis reflects the
fact that, as noted above,  that term is expected to behave
as the energy, and hence to have exponent
$\alpha - 1$, where $\alpha = {{1}\over{2}}$.
From the above, it follows that the asymptotic form of the
coefficients, $c_n$, behaves like:
\begin{equation}
t_n \sim \lambda^n[a_1n^{11/32}  + b_1n^{11/32-\Delta }
	+ (-1)^nd_1n^{-3/2}] \label{ampest}
\end{equation}

The three amplitudes, $a_1,  b_1, d_1$  come from the leading singularity
, the correction-to-scaling term
and the term on the negative real axis respectively. A small program
written in Mathematica was used to fit successive triples of coefficients,
$ c_{n-2}, c_{n-1}$ and $c_n$ for $n$ = 6,7,8,...,31. The results
  (with $\Delta = {1 \over 2}$) are shown in Table \ref{amp1}

At first sight, these appear to be converging rather well.
Closer inspection reveals
that the sequences have a turning point at around n=29. We next tried a higher
value of $\Delta$, choosing $\Delta = 0.75$ in agreement with the prediction
of the transformation method of Adler et al. cited above.
The results are shown in Table  \ref{amp2}

These sequences of
amplitudes appear to be converging reasonably well, and support the
earlier finding that the correction-to-scaling exponent is around $0.75$.
If this is correct, we can extrapolate the above sequences and find
$a_1 = 1.272 \pm 0.002$, $b_1 = -0.32 \pm 0.02$ and $d_1 = 0.035 \pm 0.004$.
Even if the correction to scaling exponent were not as estimated, the
leading amplitude is still likely to be within the quoted range.

%
%
%
%
%
%
%
%
%
%
%

\acknowledgements

We would like to thank Ian G Enting for introducing us to
the finite lattice method.
One of us (A.R.C.) would like to thank the A.O. Capell, Wyselaskie
and Daniel Curdie scholarships.
The other (A.J.G.) would like to thank the ARC for financial support.

\newpage
\bibliographystyle{unsrt}
\bibliography{comb}

\idrevtex{
\newpage
}

\figure{
Two examples of a partial path, each with
	a boundary (vertical line) and the same boundary conditions.
\\
\input{f2}
\label{exampbound}
}

\figure{
Two boundaries differing by only one site.
\\
\input{f3}
\label{twobound}
}

\figure{
The two new partial
	paths resulting from moving
	the active boundary across one site. A third possibility is not to
	add any bonds and have two floating
	partial paths instead of one floating
	partial path and two connected path ends.
\\
\input{f4}
\label{exampnew}
}
\newpage

\begin{table}
\caption{Number of boundary conditions for trails ($B_n$) and
	 for SAWs ($s_n$). Values for $s_n$ come from\cite{saw:saw39}.
	 }
\begin{tabular}{cccc}
$n$ & $L_n$ & $B_n$ & $s_n$           \\
\tableline
-1 & 1 & 2 & 2 \\
0 & 2 & 5 & 5 \\
1 & 4 & 13 & 13\\
2 & 10 & 38 & 37 \\
3 & 26 & 116 & 106 \\
4 & 76 & 382 & 312\\
5 & 232 & 1310 & 925\\
6 & 764 & 4748 & 2767 \\
7 & 2620 & 17848 & 8314\\
8 & 9496 & 70076 & 25073\\
9 & 35696 & 284252 & 75791\\
10 & 140152 & 1195240 & 229495\\
\end{tabular}
\label{BLtable}
\end{table}

\begin{table}
\caption{ Numbers of trails $t_n$ of $n$ steps }
\begin{tabular}{rl}
$n$ & $t_n$ \\
\tableline
0 & 1 \\
1 & 4 \\
2 & 12 \\
3 & 36 \\
4 & 108 \\
5 & 316 \\
6 & 916 \\
7 & 2628 \\
8 & 7500 \\
9 & 21268 \\
10 & 60092 \\
11 & 169092 \\
12 & 474924 \\
13 & 1329188 \\
14 & 3715244 \\
15 & 10359636 \\
16 & 28856252 \\
17 & 80220244 \\
18 & 222847804 \\
19 & 618083972 \\
20 & 1713283628 \\
21 & 4742946484 \\
22 & 13123882524 \\
23 & 36274940740 \\
24 & 100226653420 \\
25 & 276669062116 \\
26 & 763482430316 \\
27 & 2105208491748 \\
28 & 5803285527724 \\
29 & 15986580203460 \\
30 & 44028855864492 \\
31 & 121187822490084 \\
\end{tabular}
\label{tabenum}
\end{table}

\begin{table}
\caption{
Sequences of amplitude estimates assuming $\Delta = {1 \over 2}$
Refer equation (\ref{ampest}) }
\begin{tabular}{cccc}
       $n$ & $d_1$ & $b_1$ & $a_1$ \\
\tableline
21 & 0.0289 & -0.1805 & 1.2795\\
22 & 0.0309 & -0.1833 & 1.2801\\
23 & 0.0296 & -0.1849 & 1.2805\\
24 & 0.0311 & -0.1868 & 1.2809\\
25 & 0.0306 & -0.1874 & 1.2810\\
26 & 0.0313 & -0.1885 & 1.2812\\
27 & 0.0310 & -0.1889 & 1.2813\\
28 & 0.0314 & -0.1894 & 1.2814\\
29 & 0.0315 & -0.1893 & 1.2814\\
30 & 0.0316 & -0.1894 & 1.2814\\
31 & 0.0319 & -0.1890 & 1.2813\\
\end{tabular}
\label{amp1}
\end{table}

\begin{table}
\caption{
Sequences of amplitude estimates assuming $\Delta = {3 \over 4}$
Refer equation (\ref{ampest}) }
\begin{tabular}{cccc}
       $n$ & $d_1$ & $b_1$ & $a_1$ \\
\tableline
21 & 0.0281 & -0.2544 & 1.2661\\
22 & 0.0316 & -0.2615 & 1.2668\\
23 & 0.0289 & -0.2670 & 1.2673	\\
24 & 0.0318 & -0.2727 & 1.2679\\
25 & 0.0298 & -0.2765 & 1.2683\\
26 & 0.0321 & -0.2809 & 1.2687\\
27 & 0.0303 & -0.2878 & 1.2690\\
28 & 0.0321 & -0.2902 & 1.2692\\
29 & 0.0309 & -0.2902 & 1.2695\\
30 & 0.0323 & -0.2930 & 1.2697\\
31 & 0.0313 & -0.2949 & 1.2698\\
\end{tabular}
\label{amp2}
\end{table}

\end{document}

%% file: P.tex
\setlength{\unitlength}{0.0025in}%
\begin{picture}(280,40)(120,200)
\thinlines
\put(260,220){\circle{40}}
\put(140,220){\circle{40}}
\put(380,220){\circle{40}}
\thicklines
\put(280,220){\line( 1, 0){ 80}}
\put(160,220){\line( 1, 0){ 80}}
\end{picture}

%% file: Q.tex
\setlength{\unitlength}{0.0025in}%
\begin{picture}(280,280)(240,200)
\thinlines
\put(260,340){\circle{40}}
\put(380,460){\circle{40}}
\put(380,340){\circle{40}}
\put(380,220){\circle{40}}
\put(500,340){\circle{40}}
\put(260,460){\circle{40}}
\put(500,220){\circle{40}}
\thicklines
\put(280,460){\line( 1, 0){ 80}}
\put(260,360){\line( 0, 1){ 80}}
\put(380,440){\line( 0,-1){ 80}}
\put(380,320){\line( 0,-1){ 80}}
\put(400,220){\line( 1, 0){ 80}}
\put(500,240){\line( 0, 1){ 80}}
\end{picture}

%% file: R.tex
\setlength{\unitlength}{0.0025in}%
\begin{picture}(160,280)(360,200)
\thicklines
\put(500,360){\line( 0, 1){ 80}}
\thinlines
\put(380,460){\circle{40}}
\put(380,340){\circle{40}}
\put(380,220){\circle{40}}
\put(500,460){\circle{40}}
\put(500,340){\circle{40}}
\thicklines
\put(500,240){\line( 0, 1){ 80}}
\thinlines
\put(500,220){\circle{40}}
\thicklines
\put(380,440){\line( 0,-1){ 80}}
\put(380,320){\line( 0,-1){ 80}}
\put(400,220){\line( 1, 0){ 80}}
\end{picture}

%% file: S.tex
\setlength{\unitlength}{0.0025in}%
\begin{picture}(280,280)(360,200)
\thicklines
\put(500,360){\line( 0, 1){ 80}}
\thinlines
\put(500,220){\circle{40}}
\thicklines
\put(400,220){\line( 1, 0){ 80}}
\thinlines
\put(620,460){\circle{40}}
\thicklines
\put(520,460){\line( 1, 0){ 80}}
\thinlines
\put(380,340){\circle{40}}
\put(500,460){\circle{40}}
\thicklines
\put(380,440){\line( 0,-1){ 80}}
\put(500,240){\line( 0, 1){ 80}}
\thinlines
\put(620,340){\circle{40}}
\thicklines
\put(620,440){\line( 0,-1){ 80}}
\thinlines
\put(380,460){\circle{40}}
\put(380,220){\circle{40}}
\put(500,340){\circle{40}}
\thicklines
\put(380,320){\line( 0,-1){ 80}}
\end{picture}

%% file: T.tex
\setlength{\unitlength}{0.0025in}%
\begin{picture}(280,280)(360,200)
\thicklines
\put(500,360){\line( 0, 1){ 80}}
\thinlines
\put(500,220){\circle{40}}
\thicklines
\put(400,220){\line( 1, 0){ 80}}
\thinlines
\put(620,460){\circle{40}}
\thicklines
\put(520,460){\line( 1, 0){ 80}}
\thinlines
\put(380,340){\circle{40}}
\thicklines
\put(380,440){\line( 0,-1){ 80}}
\thinlines
\put(500,460){\circle{40}}
\put(620,340){\circle{40}}
\thicklines
\put(500,240){\line( 0, 1){ 80}}
\put(620,440){\line( 0,-1){ 80}}
\thinlines
\put(620,220){\circle{40}}
\thicklines
\put(620,320){\line( 0,-1){ 80}}
\thinlines
\put(380,460){\circle{40}}
\put(380,220){\circle{40}}
\put(500,340){\circle{40}}
\thicklines
\put(380,320){\line( 0,-1){ 80}}
\end{picture}

%% file: V.tex
\setlength{\unitlength}{0.0025in}%
\begin{picture}(400,700)(240,80)
\thicklines
\put(620,120){\vector( 0, 1){ 80}}
\thinlines
\put(380,220){\circle{40}}
\put(500,220){\circle{40}}
\thicklines
\put(380,240){\vector( 0, 1){ 80}}
\put(500,320){\vector( 0,-1){ 80}}
\put(480,220){\vector(-1, 0){ 80}}
\thinlines
\put(380,460){\circle{40}}
\put(500,340){\circle{40}}
\put(380,340){\circle{40}}
\put(500,460){\circle{40}}
\put(620,460){\circle{40}}
\put(620,220){\circle{40}}
\put(620,340){\circle{40}}
\thicklines
\put(360,580){\vector(-1, 0){ 80}}
\thinlines
\put(620,100){\circle{40}}
\put(260,580){\circle{40}}
\put(380,580){\circle{40}}
\put(260,700){\circle{40}}
\thicklines
\put(380,360){\vector( 0, 1){ 80}}
\put(500,440){\vector( 0,-1){ 80}}
\put(600,460){\vector(-1, 0){ 80}}
\put(620,360){\vector( 0, 1){ 80}}
\put(620,240){\vector( 0, 1){ 80}}
\put(380,480){\vector( 0, 1){ 80}}
\put(260,600){\vector( 0, 1){ 80}}
\end{picture}

%% file: X.tex
\setlength{\unitlength}{0.0025in}%
\begin{picture}(280,280)(360,200)
\thinlines
\put(380,220){\circle{40}}
\put(500,340){\circle{40}}
\put(380,340){\circle{40}}
\put(500,460){\circle{40}}
\put(620,460){\circle{40}}
\put(620,220){\circle{40}}
\put(620,340){\circle{40}}
\put(380,460){\circle{40}}
\put(500,220){\circle{40}}
\thicklines
\put(480,220){\vector(-1, 0){ 80}}
\put(380,240){\vector( 0, 1){ 80}}
\put(500,320){\vector( 0,-1){ 80}}
\put(380,360){\vector( 0, 1){ 80}}
\put(500,440){\vector( 0,-1){ 80}}
\put(600,460){\vector(-1, 0){ 80}}
\put(620,240){\vector( 0, 1){ 80}}
\put(620,360){\vector( 0, 1){ 80}}
\end{picture}

%% file: wzX.tex
\setlength{\unitlength}{0.0025in}%
\begin{picture}(280,400)(360,80)
\thicklines
\put(620,240){\vector( 0, 1){ 80}}
\thinlines
\put(620,100){\circle{40}}
\thicklines
\put(620,120){\vector( 0, 1){ 80}}
\thinlines
\put(380,460){\circle{40}}
\put(380,220){\circle{40}}
\put(500,340){\circle{40}}
\put(380,340){\circle{40}}
\put(500,460){\circle{40}}
\put(620,220){\circle{40}}
\put(620,460){\circle{40}}
\thicklines
\put(480,220){\vector(-1, 0){ 80}}
\thinlines
\put(620,340){\circle{40}}
\put(500,220){\circle{40}}
\thicklines
\put(380,240){\vector( 0, 1){ 80}}
\put(500,320){\vector( 0,-1){ 80}}
\put(380,360){\vector( 0, 1){ 80}}
\put(500,440){\vector( 0,-1){ 80}}
\put(600,460){\vector(-1, 0){ 80}}
\put(620,360){\vector( 0, 1){ 80}}
\end{picture}

%% file: red.tex
\setlength{\unitlength}{0.0025in}%
\begin{picture}(520,760)(240,80)
\thinlines
\put(500,340){\circle{40}}
\put(380,340){\circle{40}}
\put(620,340){\circle{40}}
\put(620,100){\circle{40}}
\put(260,580){\circle{40}}
\put(380,580){\circle{40}}
\put(260,700){\circle{40}}
\put(380,220){\circle{40}}
\put(500,220){\circle{40}}
\put(620,220){\circle{40}}
\put(740,100){\circle{40}}
\put(260,820){\circle{40}}
\put(380,820){\circle{40}}
\put(620,460){\circle{40}}
\put(380,700){\circle{40}}
\put(500,700){\circle{40}}
\put(500,460){\circle{40}}
\put(380,460){\circle{40}}
\put(500,820){\circle{40}}
\thicklines
\put(500,720){\line( 0, 1){ 80}}
\put(380,440){\line( 0,-1){ 80}}
\put(620,320){\line( 0,-1){ 80}}
\put(380,560){\line( 0,-1){ 80}}
\put(260,680){\line( 0,-1){ 80}}
\put(280,580){\line( 1, 0){ 80}}
\put(380,320){\line( 0,-1){ 80}}
\put(500,240){\line( 0, 1){ 80}}
\put(400,220){\line( 1, 0){ 80}}
\put(620,200){\line( 0,-1){ 80}}
\put(640,100){\line( 1, 0){ 80}}
\put(260,720){\line( 0, 1){ 80}}
\put(380,800){\line( 0,-1){ 80}}
\put(280,820){\line( 1, 0){ 80}}
\put(620,440){\line( 0,-1){ 80}}
\put(520,460){\line( 1, 0){ 80}}
\put(400,700){\line( 1, 0){ 80}}
\put(500,360){\line( 0, 1){ 80}}
\end{picture}

%% file: f2.tex
\setlength{\unitlength}{0.006in}%
\begin{picture}(890,370)(0,450)
\thinlines
\put(750,630){\circle{20}}
\put(750,570){\circle{20}}
\put(750,750){\circle{20}}
\put(750,810){\circle{20}}
\put(750,510){\circle{20}}
\put(810,690){\circle{20}}
\put(810,630){\circle{20}}
\put(810,570){\circle{20}}
\put(810,750){\circle{20}}
\put(810,810){\circle{20}}
\put(810,510){\circle{20}}
\put(870,690){\circle{20}}
\put(870,630){\circle{20}}
\put(870,570){\circle{20}}
\put(870,750){\circle{20}}
\put(870,810){\circle{20}}
\put(870,510){\circle{20}}
\put(690,690){\circle{20}}
\put(690,630){\circle{20}}
\put(690,570){\circle{20}}
\put(690,750){\circle{20}}
\put(690,810){\circle{20}}
\put(690,510){\circle{20}}
\put(630,690){\circle{20}}
\put(630,630){\circle{20}}
\put(630,570){\circle{20}}
\put(630,750){\circle{20}}
\put(630,810){\circle{20}}
\put(630,510){\circle{20}}
\put(570,690){\circle{20}}
\put(570,630){\circle{20}}
\put(570,570){\circle{20}}
\put(570,750){\circle{20}}
\put(570,810){\circle{20}}
\put(570,510){\circle{20}}
\put(510,690){\circle{20}}
\put(510,630){\circle{20}}
\put(510,570){\circle{20}}
\put(510,750){\circle{20}}
\put(510,810){\circle{20}}
\put(510,510){\circle{20}}
\put(780,819){\line( 0,-1){319}}
\thicklines
\put(800,750){\line(-1, 0){ 40}}
\put(800,690){\line(-1, 0){ 40}}
\put(740,750){\line(-1, 0){ 40}}
\put(680,750){\line(-1, 0){ 40}}
\put(620,750){\line(-1, 0){ 40}}
\put(570,740){\line( 0,-1){ 40}}
\put(560,690){\line(-1, 0){ 40}}
\put(510,680){\makebox(0.4444,0.6667){\tenrm .}}
\put(510,680){\line( 0,-1){ 40}}
\put(510,620){\line( 0,-1){ 40}}
\put(510,560){\line( 0,-1){ 40}}
\thinlines
\put(750,690){\circle{20}}
\thicklines
\put(520,510){\line( 1, 0){ 40}}
\put(320,690){\line(-1, 0){ 40}}
\put(570,520){\line( 0, 1){ 40}}
\put(570,580){\line( 0, 1){ 40}}
\put(580,630){\line( 1, 0){ 40}}
\put(630,620){\line( 0,-1){ 40}}
\put(640,570){\line( 1, 0){ 40}}
\put(690,580){\line( 0, 1){ 40}}
\put(700,630){\line( 1, 0){ 40}}
\put(750,640){\line( 0, 1){ 40}}
\thinlines
\put(330,690){\circle{20}}
\put(330,630){\circle{20}}
\put(330,570){\circle{20}}
\put(330,750){\circle{20}}
\put(330,810){\circle{20}}
\put(330,510){\circle{20}}
\put(390,690){\circle{20}}
\put(390,630){\circle{20}}
\put(390,570){\circle{20}}
\put(390,750){\circle{20}}
\put(390,810){\circle{20}}
\put(390,510){\circle{20}}
\put(210,690){\circle{20}}
\put(210,630){\circle{20}}
\put(210,570){\circle{20}}
\put(210,750){\circle{20}}
\put(210,810){\circle{20}}
\put(210,510){\circle{20}}
\put(150,690){\circle{20}}
\put(150,630){\circle{20}}
\put(150,570){\circle{20}}
\put(150,750){\circle{20}}
\put(150,810){\circle{20}}
\put(150,510){\circle{20}}
\put( 90,690){\circle{20}}
\put( 90,630){\circle{20}}
\put( 90,570){\circle{20}}
\put( 90,750){\circle{20}}
\put( 90,810){\circle{20}}
\put( 90,510){\circle{20}}
\put( 30,690){\circle{20}}
\put( 30,630){\circle{20}}
\put( 30,570){\circle{20}}
\put( 30,750){\circle{20}}
\put( 30,810){\circle{20}}
\put( 30,510){\circle{20}}
\thicklines
\put(270,700){\line( 0, 1){ 40}}
\thinlines
\put(270,690){\circle{20}}
\put(270,630){\circle{20}}
\put(270,570){\circle{20}}
\put(270,750){\circle{20}}
\put(270,810){\circle{20}}
\put(270,510){\circle{20}}
\put(300,819){\line( 0,-1){319}}
\thicklines
\put(320,750){\line(-1, 0){ 40}}
\end{picture}

%% file: f3.tex
\setlength{\unitlength}{0.006in}%
\begin{picture}(760,640)(20,180)
\thinlines
\put(520,440){\circle{40}}
\put(520,320){\circle{40}}
\put(520,680){\circle{40}}
\put(520,800){\circle{40}}
\put(520,200){\circle{40}}
\put(640,560){\circle{40}}
\put(640,440){\circle{40}}
\put(640,320){\circle{40}}
\put(640,680){\circle{40}}
\put(640,800){\circle{40}}
\put(640,200){\circle{40}}
\put(760,560){\circle{40}}
\put(760,440){\circle{40}}
\put(760,320){\circle{40}}
\put(760,680){\circle{40}}
\put(760,800){\circle{40}}
\put(760,200){\circle{40}}
\put(400,560){\circle{40}}
\put(400,440){\circle{40}}
\put(400,320){\circle{40}}
\put(400,680){\circle{40}}
\put(400,800){\circle{40}}
\put(400,200){\circle{40}}
\put(280,560){\circle{40}}
\put(280,440){\circle{40}}
\put(280,320){\circle{40}}
\put(280,680){\circle{40}}
\put(280,800){\circle{40}}
\put(280,200){\circle{40}}
\put(160,560){\circle{40}}
\put(160,440){\circle{40}}
\put(160,320){\circle{40}}
\put(160,680){\circle{40}}
\put(520,560){\circle{40}}
\put(160,800){\circle{40}}
\put(600,180){\line( 0, 1){560}}
\put(600,740){\line( 1, 0){100}}
\put(700,740){\line( 0, 1){ 80}}
\put(160,200){\circle{40}}
\put( 40,560){\circle{40}}
\put( 40,440){\circle{40}}
\put( 40,320){\circle{40}}
\put( 40,680){\circle{40}}
\put( 40,800){\circle{40}}
\put( 40,200){\circle{40}}
\thicklines
\put( 60,800){\line( 1, 0){ 80}}
\put(180,800){\line( 1, 0){ 80}}
\put(280,780){\line( 0,-1){ 80}}
\put(300,680){\line( 1, 0){ 80}}
\put(400,700){\line( 0, 1){ 80}}
\put(420,800){\line( 1, 0){ 80}}
\put(540,800){\line( 1, 0){ 80}}
\put( 40,780){\line( 0,-1){ 80}}
\put( 40,660){\line( 0,-1){ 80}}
\put( 40,540){\line( 0,-1){ 80}}
\put( 60,440){\line( 1, 0){ 80}}
\put(160,460){\line( 0, 1){ 80}}
\put(180,560){\line( 1, 0){ 80}}
\put(280,540){\line( 0,-1){ 80}}
\put(280,420){\line( 0,-1){ 80}}
\put(160,300){\line( 0,-1){ 80}}
\put(180,200){\line( 1, 0){ 80}}
\put(300,200){\line( 1, 0){ 80}}
\put(420,200){\line( 1, 0){ 80}}
\put(540,200){\line( 1, 0){ 80}}
\put(300,320){\line( 1, 0){ 80}}
\put(420,320){\line( 1, 0){ 80}}
\put(520,340){\line( 0, 1){ 80}}
\put(540,440){\line( 1, 0){ 80}}
\thinlines
\put(560,820){\line( 0,-1){640}}
\end{picture}

%% file: f4.tex
\setlength{\unitlength}{0.006in}%
\begin{picture}(880,320)(20,500)
\thinlines
\put(270,630){\circle{20}}
\put(270,570){\circle{20}}
\put(270,750){\circle{20}}
\put(270,810){\circle{20}}
\put(270,510){\circle{20}}
\put(330,690){\circle{20}}
\put(330,630){\circle{20}}
\put(330,570){\circle{20}}
\put(330,750){\circle{20}}
\put(330,810){\circle{20}}
\put(330,510){\circle{20}}
\put(390,690){\circle{20}}
\put(390,630){\circle{20}}
\put(390,570){\circle{20}}
\put(390,750){\circle{20}}
\put(390,810){\circle{20}}
\put(390,510){\circle{20}}
\put(210,690){\circle{20}}
\put(210,630){\circle{20}}
\put(210,570){\circle{20}}
\put(210,750){\circle{20}}
\put(210,810){\circle{20}}
\put(210,510){\circle{20}}
\put(150,690){\circle{20}}
\put(150,630){\circle{20}}
\put(150,570){\circle{20}}
\put(150,750){\circle{20}}
\put(150,810){\circle{20}}
\put(150,510){\circle{20}}
\put( 90,690){\circle{20}}
\put( 90,630){\circle{20}}
\put( 90,570){\circle{20}}
\put( 90,750){\circle{20}}
\put( 90,810){\circle{20}}
\put( 90,510){\circle{20}}
\put( 30,690){\circle{20}}
\put( 30,630){\circle{20}}
\put( 30,570){\circle{20}}
\put( 30,750){\circle{20}}
\put( 30,810){\circle{20}}
\put( 30,510){\circle{20}}
\put(770,690){\circle{20}}
\put(770,630){\circle{20}}
\put(770,570){\circle{20}}
\put(770,750){\circle{20}}
\put(770,810){\circle{20}}
\put(770,510){\circle{20}}
\put(830,690){\circle{20}}
\put(830,630){\circle{20}}
\put(830,570){\circle{20}}
\put(830,750){\circle{20}}
\put(830,810){\circle{20}}
\put(830,510){\circle{20}}
\put(890,690){\circle{20}}
\put(890,630){\circle{20}}
\put(890,570){\circle{20}}
\put(890,750){\circle{20}}
\put(890,810){\circle{20}}
\put(890,510){\circle{20}}
\put(710,690){\circle{20}}
\put(710,630){\circle{20}}
\put(710,570){\circle{20}}
\put(710,750){\circle{20}}
\put(710,810){\circle{20}}
\put(710,510){\circle{20}}
\put(650,690){\circle{20}}
\put(650,630){\circle{20}}
\put(650,570){\circle{20}}
\put(650,750){\circle{20}}
\put(650,810){\circle{20}}
\put(650,510){\circle{20}}
\put(590,690){\circle{20}}
\put(590,630){\circle{20}}
\put(590,570){\circle{20}}
\put(270,690){\circle{20}}
\put(590,750){\circle{20}}
\put(810,500){\line( 0, 1){280}}
\put(810,780){\line( 1, 0){ 50}}
\put(860,780){\line( 0, 1){ 40}}
\put(590,810){\circle{20}}
\put(590,510){\circle{20}}
\put(530,690){\circle{20}}
\put(530,630){\circle{20}}
\put(530,570){\circle{20}}
\put(530,750){\circle{20}}
\put(530,810){\circle{20}}
\put(530,510){\circle{20}}
\thicklines
\put(150,800){\line( 0,-1){ 40}}
\put(330,800){\line( 0,-1){ 40}}
\thinlines
\put(210,690){\circle{20}}
\put(210,630){\circle{20}}
\put(210,570){\circle{20}}
\put(210,750){\circle{20}}
\put(210,810){\circle{20}}
\put(210,510){\circle{20}}
\put(210,690){\circle{20}}
\put(210,630){\circle{20}}
\put(210,570){\circle{20}}
\put(210,750){\circle{20}}
\put(210,810){\circle{20}}
\put(210,510){\circle{20}}
\thicklines
\put(600,690){\line( 1, 0){ 40}}
\put(840,810){\line( 1, 0){ 40}}
\put( 40,810){\line( 1, 0){ 40}}
\put(100,810){\line( 1, 0){ 40}}
\put(160,750){\line( 1, 0){ 40}}
\put(210,760){\line( 0, 1){ 40}}
\put(220,810){\line( 1, 0){ 40}}
\put(280,810){\line( 1, 0){ 40}}
\put( 30,800){\line( 0,-1){ 40}}
\put( 30,740){\line( 0,-1){ 40}}
\put( 30,680){\line( 0,-1){ 40}}
\put( 40,630){\line( 1, 0){ 40}}
\put( 90,640){\line( 0, 1){ 40}}
\put(100,690){\line( 1, 0){ 40}}
\put(150,680){\line( 0,-1){ 40}}
\put(150,620){\line( 0,-1){ 40}}
\put( 90,560){\line( 0,-1){ 40}}
\put(100,510){\line( 1, 0){ 40}}
\put(160,510){\line( 1, 0){ 40}}
\put(220,510){\line( 1, 0){ 40}}
\put(280,510){\line( 1, 0){ 40}}
\put(160,570){\line( 1, 0){ 40}}
\put(220,570){\line( 1, 0){ 40}}
\put(270,580){\line( 0, 1){ 40}}
\put(280,630){\line( 1, 0){ 40}}
\thinlines
\put(290,820){\line( 0,-1){320}}
\put(310,500){\line( 0, 1){280}}
\put(310,780){\line( 1, 0){ 50}}
\put(360,780){\line( 0, 1){ 40}}
\thicklines
\put(540,810){\line( 1, 0){ 40}}
\put(600,810){\line( 1, 0){ 40}}
\put(650,800){\line( 0,-1){ 40}}
\put(660,750){\line( 1, 0){ 40}}
\put(710,760){\line( 0, 1){ 40}}
\put(720,810){\line( 1, 0){ 40}}
\put(780,810){\line( 1, 0){ 40}}
\put(530,800){\line( 0,-1){ 40}}
\put(530,740){\line( 0,-1){ 40}}
\put(530,680){\line( 0,-1){ 40}}
\put(540,630){\line( 1, 0){ 40}}
\put(590,640){\line( 0, 1){ 40}}
\put(650,680){\line( 0,-1){ 40}}
\put(650,620){\line( 0,-1){ 40}}
\put(590,560){\line( 0,-1){ 40}}
\put(600,510){\line( 1, 0){ 40}}
\put(660,510){\line( 1, 0){ 40}}
\put(720,510){\line( 1, 0){ 40}}
\put(780,510){\line( 1, 0){ 40}}
\put(660,570){\line( 1, 0){ 40}}
\put(720,570){\line( 1, 0){ 40}}
\put(770,580){\line( 0, 1){ 40}}
\put(780,630){\line( 1, 0){ 40}}
\thinlines
\put(790,820){\line( 0,-1){320}}
\end{picture}